# BATCH FABRICATION OF CLUSTER ASSEMBLED MICROARRAYS FOR CHEMICAL SENSING

*E. Barborini[1], M. Leccardi[1], G. Bertolini[1], O. Rorato[1], P. Repetto[1], D. Bandiera[1], M. Gatelli[1], A. Raso[2], A. Garibbo[2], L. Seminara[2], C. Ducati[3], P. Milani[1,4,5]*

[1]Tethis srl, via Russoli 3, 20151 Milano, Italy
[2]Selex-comms, via Negrone 1/A, 16153 Genova, Italy
[3]Materials Science and Metallurgy, University of Cambridge, Pembroke Street, Cambridge CB2 3QZ, UK
[4]Dipartimento di Fisica, Università di Milano, via Celoria 16, 20133 Milano, Italy
[5]CIMAINA, via Celoria 16, 20133 Milano, Italy

## ABSTRACT

Deposition of clusters from the gas phase is becoming an enabling technology for the production of nanostructured devices. Supersonic clusters beam deposition (SCBD) has been shown as a viable route for the production of nanostructured thin films. By using SCBD and by exploiting aerodynamical effects typical of supersonic beams it is possible to obtain very high deposition rates with a control on neutral cluster mass distribution, allowing the deposition of thin films with tailored nanostructure. Due to high deposition rates, high lateral resolution, low temperature processing, SCBD can be used for the integration of cluster-assembled films on micro- and nanofabricated platforms with limited or no post-growth processing. Here we present the industrial opportunities for batch fabrication of gas sensor microarrays based on transition metal oxide nanoparticles deposited on microfabricated substrates.

## 1. INTRODUCTION

At high temperature the chemical composition of the surface of semiconductor metal oxides is in equilibrium with the composition of the atmosphere to which it is exposed. If atmosphere change occurs, alteration of the surface chemistry also happens, as a consequence. The electrical conductivity of the oxide surface follows these modifications. This phenomenon is exploited in conductimetric gas sensors [1]. The most important materials for conductimetric gas sensors are simple metal oxides, such as $SnO_2$, $TiO_2$, $WO_3$, $In_2O_3$, ZnO.

The use of nanostructured thin films of oxides as sensing layer has recently attracted conspicuous interest due to their huge specific surface increasing the interaction with gas-phase species. Several methods can be employed to produce nanostructured oxide films, from sol-gel and wet chemistry processes in general, to physical methods such as sputtering or evaporation.

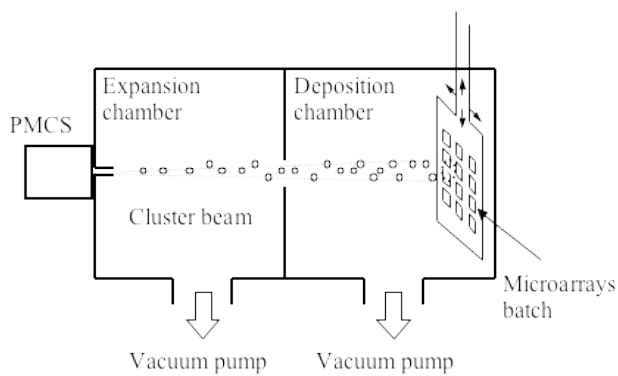

**Figure 1**. Schematic of the SCBD unit with nanoparticle source (PMCS), beam formation zone (expansion chamber), and deposition chamber equipped with hard mask system for batch deposition of microarrays.

Among various deposition techniques, Supersonic Cluster Beam Deposition (SCBD) appeared to be very promising as nanoparticles can be directly deposited on micro-machined platforms with limited or no post-growth processing [2, 3].

Low impact energy (soft landing) and limited diffusion are characteristics of SCBD process, causing the film to grow accordingly to a highly porous structure at the nanoscales. Moreover, SCBD offers the possibility to deposit nanostructured films on every kind of substrates, silicon membranes included, at room temperature, in ultra-clean conditions. Due to the high collimation of the cluster beam, patterned depositions with micrometric

*E. Barborini. M. Leccardi, G. Bertolini, O. Rorato, P. Repetto, D. Bandiera,*
*M. Gatelli, A. Raso, A. Garibbo, L. Seminara, C. Ducati, P. Milani*

# Batch Fabrication of Cluster Assembled Microarrays for Chemical Sensing

resolution can be obtained by hard mask method [4]. This permits parallel deposition of identical devices (batch) avoiding the use of photoresist and chemical etching, typical of photolithographic process, that could contaminate the platform/substrate surface.

Here, we report on batch deposition of chemical sensors, based on heterogeneous microarrays of nanostructured oxides thin films, by SCBD. We also report on their sensing properties respect to chemical species related to environmental pollution, and to volatile organic compounds (VOC), such as ethanol.

## 2. EXPERIMENTAL

Nanostructured films of TiO$_2$, WO$_3$, SnO$_2$, and MoO$_3$, having thickness around some hundreds of nanometers, were produced by SCBD. The cluster beam was generated by a Pulsed Microplasma Cluster Source (PMCS) [5]. The operation principle of the PMCS is based on the ablation of the metallic target by a plasma jet of inert gas (typically Argon) ignited by a pulsed electric discharge. After ablation, metallic atoms thermalize into the inert gas and re-condense to form clusters that are entrained by the gas flux towards PMCS exit nozzle and extracted by nozzle expansion. A set of aerodynamic lenses [6] collects the gas-nanoparticles stream from PMCS nozzle and forces the nanoparticles to concentrate close to beam axis, increasing the beam collimation (divergence less than 20 mrad), and the in-axis intensity. The beam collimation allows the separation of the deposition chamber from the nozzle expansion chamber, with differential vacuum approach, in order to reach ultra-clean conditions in deposition chamber. Moreover, various PMCS beamlines can face the same deposition chamber for co-deposition or multi-layer deposition. The deposition chamber is equipped with a motorized manipulator to allow large area depositions (50×220 mm$^2$) by sample holder rastering. Fig. 1 schematically shows the structure of the deposition apparatus.

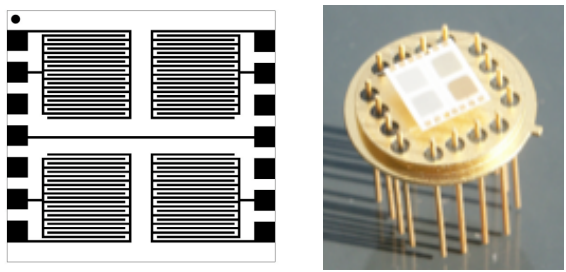

**Figure 2**. Left: layout of the micro-machined platforms with array structure. On front side, beside the four pairs of interdigitized metallizations, a Pt thin-wire thermometer was integrated, while on back side a thin film heater is present. Right: the microarray assembled into a standard 16 pin TO package.

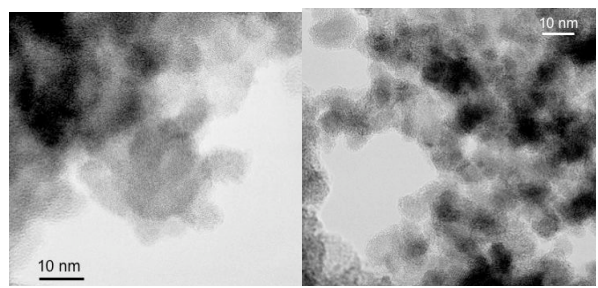

**Figure 3**. TEM image of as-deposited tungsten oxide sample (left). The film is composed by amorphous nanoparticles having an average size of 10 nm, assembled in a porous structure. The same material (right) after annealing at 200°C. Lattice fringes appear, indicating the polycrystalline nature of nanoparticles, while average size and porous structure remain very similar to those of as deposited material.

By exploiting hard mask patterning, we developed micro-machined platforms having an array structure in order to deposit different oxides on each single element of the array (heterogeneous microarray). A thin film heater and a Pt thin-wire thermometer were integrated on the substrate to control the operation temperature of the sensor. Fig. 2 shows the layout of the microarray platform. The sample holder into the deposition chamber was developed in order to host up to 25 microarray platforms for batch deposition.

The proper oxide stoichiometry is reached during post-deposition high temperature annealing in air (400°C). Besides stoichiometry adjustment, annealing is needed to fix the nanostructure of the sensing materials in order to avoid any further modification during sensor operation.

## 3. RESULTS AND DISCUSSION

The as deposited films generally have an amorphous and porous structure at the nanoscales, attributed to particle impact with low kinetic energy and limited diffusion. After annealing, the amorphous grains rearrange into a crystalline structure. For example, Fig. 3 shows Transmission Electron Microscopy (TEM) images of as deposited and 200°C annealed tungsten oxide films. The as deposited material is composed by nanoparticles having an average size of 10 nm, assembled in a porous structure. No evidence of lattice fringes is visible inside nanoparticles indicating amorphous structure. Lattice fringes are discernible into the annealed material, indicating the polycrystalline nature of nanoparticles, while average size and porous structure look very similar to those of as deposited samples. It has to be emphasized that high temperature annealing causes small grain growth and coalescence, thus the porous structure is preserved. This is crucial for all the applications where devices operate at high temperature, such as conductimetric gas sensors.

The sensing properties of the films were evaluated with respect to various species related to environmental pollution, such as CO, NO, $NO_2$, and $SO_2$, as well as volatile organic compounds (VOC), such as ethanol. By means of an automatic mixing system based on mass flow controllers, these compounds were added at trace level to pure dry air flowing into a test chamber. A simple, low-cost, front-end electronic was specifically developed to acquire the signals from the microarray sensor during the test sequence. Fig. 4 shows, as examples, the response of nanostructured $TiO_2$ film to ethanol. In this case Pd nanoparticles were added to $TiO_2$, in a two-step multi-layer deposition process involving two PMCS beamlines, to exploit the catalytic action of Pd. Base-line stability, fast response and recovery, high sensitivity, role of the operation temperature, effect of humidity, are the general issues to be addressed.

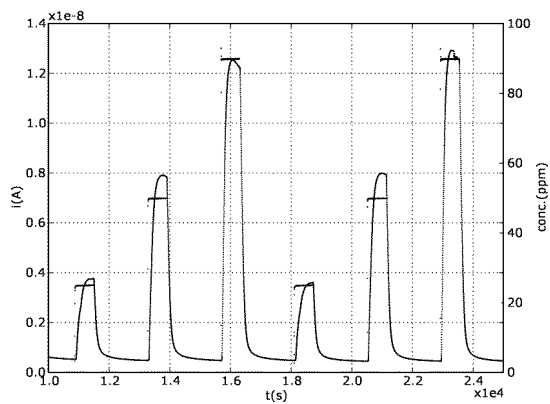

**Figure 4**. Response of *ns*-$TiO_2$ + *ns*-Pd at 300°C to ethanol. The step-shaped curve indicates the real concentration of the compound under analysis, referred to right axis.

## 4. CONCLUSIONS

Highly porous nanostructured metal oxide films for gas sensing applications can be prepared by SCBD. By hard mask patterning, it is possible to prepare in batch devices having a microarray structure with different oxides as sensing elements (heterogeneous array). Batch deposition of up to 25 identical devices was performed with the present experimental set-up. The characterization of the gas sensing properties of each single nanostructured metal oxide composing the microarray shows good results respect to species related to environmental pollution and VOC.

The use of a combinatorial approach in microarray fabrication integrated with **neural network analysis** for data handling can be of great help for the understanding of the mechanisms underlying gas selectivity and for the efficient and inexpensive realization of microsensor arrays, e.g. for environmental monitoring.

## 5. REFERENCES

[1] T. Seiyama, A. Kato, K. Fujiishi, and N. Nagatani, "A new detector for gaseous components using semiconductive thin films" *Anal. Chem.*, 34, 1502 (1962).

[2] E. Barborini, I.N. Kholmanov, P. Piseri, C. Ducati, C.E. Bottani and P. Milani, "Engineering the nanocrystalline structure of $TiO_2$ films by aerodynamically filtered cluster deposition" *Appl. Phys. Lett.*, 81, 3052 (2002).

[3] I.N. Kholmanov, E. Barborini, S. Vinati, P. Piseri, A. Podesta', C. Ducati, C. Lenardi, and P. Milani, "The influence of the precursor clusters on the structural and morphological evolution of nanostructured $TiO_2$ under thermal annealing" *Nanotech.*, 14, 1168 (2003).

[4] E. Barborini, P. Piseri, A. Podestà, P. Milani, "Cluster beam microfabrication of patterns of three-dimensional nanostructured objects" *Appl. Phys. Lett.*, 77, 1059 (2000).

[5] E. Barborini, P. Piseri and P. Milani, "A pulsed microplasma source of high intensity supersonic carbon cluster beams" *J. Physics D: Appl. Phys.*, 32, L105 (1999).

[6] P. Liu, P.J. Ziemann, D.B. Kittelson and P.H. McMurray, "Generating particle beams of controlled dimensions and divergence" *Aerosol Sci. Technol.*, 22, 293 (1995).